\documentclass[twocolumn,showpacs,prc,aps,floats]{revtex4}
\usepackage{graphicx}
\newcommand{\ds }{\displaystyle}
\newcommand{\ra}{\rightarrow}
\newcommand{\be}{\begin{equation}}
\newcommand{\ee}{\end{equation}}
\newcommand{\bea}{\begin{eqnarray}}
\newcommand{\eea}{\end{eqnarray}}
\newcommand{\ci}{\cite}
\newcommand{\bi}{\bibitem}
\newcommand{\nono}{\nonumber \\}

\newcommand{\dd}{\partial}

\newcommand{{\bfna}}{\mbox{\boldmath$\vec{\nabla}$}}

\def\dal{\,\lower0.3ex\vbox{\hrule\hbox{\vrule\kern2pt\vbox{\kern4pt\kern4pt}
\kern2pt\vrule}\hrule}\,}

\begin{document}

\title{ Assisted fusion}
%\vspace{1 true cm}
\author{G. K\"albermann}
\email{hope@vms.huji.ac.il}
\affiliation{Soil and Water department, 
The Robert H. Smith Faculty of Agriculture, Food and Environment 
Hebrew University, Rehovot 76100, Israel}

\begin{abstract}

A model of nuclear fusion consisting of a wave packet impinging into a well located between square one 
dimensional barriers is treated analytically. The wave function inside the well is calculated exactly for the
assisted tunneling induced by a
perturbation mimicking a constant electric field with arbitrary time dependence.
Conditions are found for the enhancement of fusion.
\end{abstract}
\maketitle

\section{\sl Introduction}

Quantum tunneling explains the transmission of particles or clusters of 
particles through regions that are energetically forbidden.
A time-honored success of the quantum tunneling model is the explanation of
$\alpha$ decay lifetimes of unstable nuclei by 
Gurney and Condon and Gamow\ci{condon, gamow}. More recent revisions of the tunneling paradigm in the nuclear decay context, 
have reaffirmed its validity.\ci{medeiros, holstein}.

In previous works we investigated the tunneling of a one dimensional 
metastable state between barriers, excited by a time dependent
potential.\ci{g1},\ci{g2}

In ref.\ci{g2} we found analytical expressions for the assisted tunneling 
processes. The acceleration of the decay
of a metastable state was found to be determined by the 
poles of the unperturbed wave functions
in the complex energy plane.

In the present work we generalize the approach of \ci{g2}, and find exact solutions to the assisted tunneling of a wave packet impinging from a region outside the range of the potentials, into a square well lying between square barriers.
This setup provides a simplified model for the fusion of nuclei.\ci{esb}

Fusion time scales are of the order of the
transit time of the nuclei past each other and differ from $\ds \alpha$ decay lifetimes markedly. 
Nuclear decay lifetimes are typically very long in
comparison to natural nuclear times. 
On this basis, it may be argued that fusion can not be affected in the same manner as found for the case of $\ds \alpha$ decay.
Nevertheless, in the presence of a long range perturbation, the tunneling process is sensible to longer time scales. 
Consequently, an external agent can affect the inter penetration the nuclei. 

However, for long times, it is not possible to follow the decay process numerically.\ci{vandijk}. Hence, 
analytical formulas are vital for nuclear fusion as well as for decay.
This is the motivation for the present effort.

In the next section we review the model and extend it the
case of fusion.
In section 3 we apply the results to the penetration of a wave packet into
a region between square barriers. By inspecting the formulas we are able identify the relevant parameter space for the enhancement of fusion.

\section{\sl Analysis of assisted tunneling }

The Schr\"odinger equation for a one dimensional system
consisting of a square well between square barriers is
\footnote{ In the following our units are 
{\sl fm} for length and time, and {\sl$ MeV~or~\ds fm^{-1}$} for energies, momenta and 
mass $\ds \gamma,~V_0,~m$.} 

\bea\label{h1}
i~\frac{\dd\Psi}{\dd t}&=& \frac{-1}{2~m}\frac{\dd^2 \Psi}{\dd x^2}-V_0\Theta(x_0-|x|)\Psi\nono
&+&\gamma~(\Theta(d-|x|)\Theta(|x|-x_0))\Psi
\eea

\noindent The inclusion of a square well between the barriers is needed to model
nuclear fusion.

The even and odd stationary states of eq.(\ref{h1}) for 
energies below the barrier strength $\ds \gamma$ are 

\bea\label{funcs}
\chi_{e,o}(x)&=&\frac{\varphi_{o,e}(k)}{\sqrt{\pi} n_{e,o}(k)}
\eea

\noindent where

\bea\label{even}
\varphi_e(k)=\left \{
\begin {array}{l}
cos(q x);|x|<x_0\\
A_1 e^{\kappa |x|}+  B_1 e^{-\kappa |x|}~;d>|x|>x_0\\
C_1 cos(k x)+sign(x) D_1 sin(k x);~|x|>d
\end{array}\right.
\eea

\noindent where $\ds \kappa$ and q are defined in eq.(\ref{ne_1}), 
and $\ds sign(x)=1,-1$ for $\ds x>0, x<0$.

\bea\label{odd}
\varphi_o(k)=\left \{
\begin {array}{l}
sin(q x);~|x|<x_0\\
sign(x)( A_2 e^{\kappa |x|}+ B_2 e^{-\kappa |x|});~d>|x|>x_0\\
sign(x) C_2 cos(k x)+ D_2 sin(k x);~|x|>d
\end{array} \right.
\eea
\noindent, and we have extracted a factor of $\ds \sqrt{\pi}$ from the 
normalizations for convenience.
The labels $\ds e,o$ refer to the even or odd character of the
wave functions

The set of even-odd functions is orthonormal and complete.\ci{bron,patil} 
The all important normalization factors of eq.(\ref{funcs}), 
and the amplitudes of eqs.(\ref{even},\ref{odd}). are shown in appendix A.

We subject the system to a time dependent spatially linear perturbation, higher powers in
 space can be dealt similarly,

\bea\label{potential}
V(x,t)=\mu~x~G(t)
\eea

\noindent with $\ds \mu$ a coupling constant. For the case of a spatially constant 
time-harmonic electric field of intensity $\ds E_0$ interacting with 
a nucleus of charge $\ds Z~|e|$ we have

\bea\label{th}
~\mu~G(t)=Z~|e|~E_0~sin(\omega t)
\eea

Applying a unitary transformation

\bea\label{unitary}
\Psi(x,t)&=&e^{-i\sigma}\Phi\nono
\sigma&=&\mu~x~\zeta+\int{\frac{\zeta^2~\mu^2}{2~m}~dt}\nono
\zeta&=&\int{G(t)~dt}
\eea
\noindent The Schr\"odinger equation(\ref{h1}) including the
perturbation of eq.(\ref{potential}) becomes

\bea\label{h2}
i~\frac{\dd\Phi}{\dd t}&=&\frac{-1}{2~m}\frac{\dd^2 \Phi}{\dd x^2}+
\gamma~(\Theta(x_0+d-|x|)\Theta(|x|-x_0))\Phi\nono
&-&V_0\Theta(x_0-|x|)\Phi+i\zeta(t)~\frac{\mu}{m}\frac{\dd \Phi}{\dd x}
\eea

Eq.(\ref{h2}) is solved by
expanding the wave function in the complete set of even and odd states 
of the unperturbed Schr\"odinger equation (\ref{funcs}) 

\bea\label{expansion}
\Phi(x,t)=\sum_{i=e,o}\int_0^{\infty}\chi_i(k,x)~c_i(k,t)~e^{
\frac{-i~k^2~t}{2~m}}~dk
\eea

\noindent Taking advantage of the superposition 
integral evaluated in the appendix A of \ci{g2}, adapted now to the case 
of a potential well between the barriers

\bea\label{superp}
I&=&\int{e^{-\frac{i(k'^2-k^2) t}{2m}}
\chi_e(k,x)\frac{\dd \chi_o(k',x)}{\dd x} dx}\nono
&=&\frac{q}{n_e(k) n_o(k)}\delta(k-k')
\eea
\noindent with $\ds q$ defined in eq.(\ref{ne_1}) of Appendix A,
the time evolution of the amplitudes $\ds c_{e,o}$ of eq.(\ref{expansion}) becomes

\bea\label{sch1}
\dot{c}_o(k)&=&-\frac{\mu~q~\zeta}{m~n_e(k)~n_o(k)}~c_e(k)\nono
\dot{c}_e(k)&=&\frac{\mu~q~\zeta}{m~n_e(k)~n_o(k)}~c_o(k)
\eea

\noindent a dot denoting a time derivative.

For an initial ($\ds t=0$), wave packet representing the incident nucleus 
carrying mean momentum $\ds q_0$ and centered at $\ds r_0$, 
$\ds \Psi_{q_0,r_0}(x,t=0)$, the solution of eq.(\ref{sch1}) reads

\bea\label{ampl}
c_e(k,t)&=&A_e~cos(z)+A_o~sin(z)\nono
c_o(k,t)&=&A_o~cos(z)-A_e~sin(z)\nono
z&=&\frac{\mu~q~Y}{m~n_e~n_o}\nono
Y(t)&=&\int_0^t{\zeta(t')~dt'}\nono
\eea

\noindent where 
\bea\label{aeao}
A_e&=&\frac{{\bf A}_e}{n_e}=\int_{-\infty}^{\infty}\Psi_{q_0,r_0}\chi_e~dx\nono
A_o&=&\frac{{\bf A}_o}{n_o}=\int_{-\infty}^{\infty}\Psi_{q_0,r_0}\chi_o~dx\nono
\eea

\section{\sl Assisted tunneling into a well between barriers}

 The most important contribution to
 eq.(\ref{expansion}) in the region of the well is due to the zeros of the normalization 
 factors $\ds n_{e,o}$ of eq.(\ref{funcs}) displayed in eqs.(\ref{neven},\ref{nodd}) of appendix A.\ci{g1},\ci{g2}

Figure 1 shows the inverse of the normalizations of 
eq.(\ref{funcs}) for $\ds m=15000 MeV, x_0=7 fm, d=12 fm, 
\gamma=10 MeV, V_0=40 MeV$, corresponding to an 
 $\ds ^{16}O$ nucleus for energies below the barrier.
The spikes in figure 1 are due to the extreme closeness of the minima of
the normalization factors to their complex zeros.

For eigenenergies above the barrier, the inverse of the amplitudes
show a smooth spectrum of peaks of order one in height and width.
These peaks will contribute a negligible amount at long times due to
the strong oscillations of the time dependent exponential factors of
the wave functions, and are henceforth omitted. 

\begin{figure}[htb]
\includegraphics[width=8cm,height=10cm]{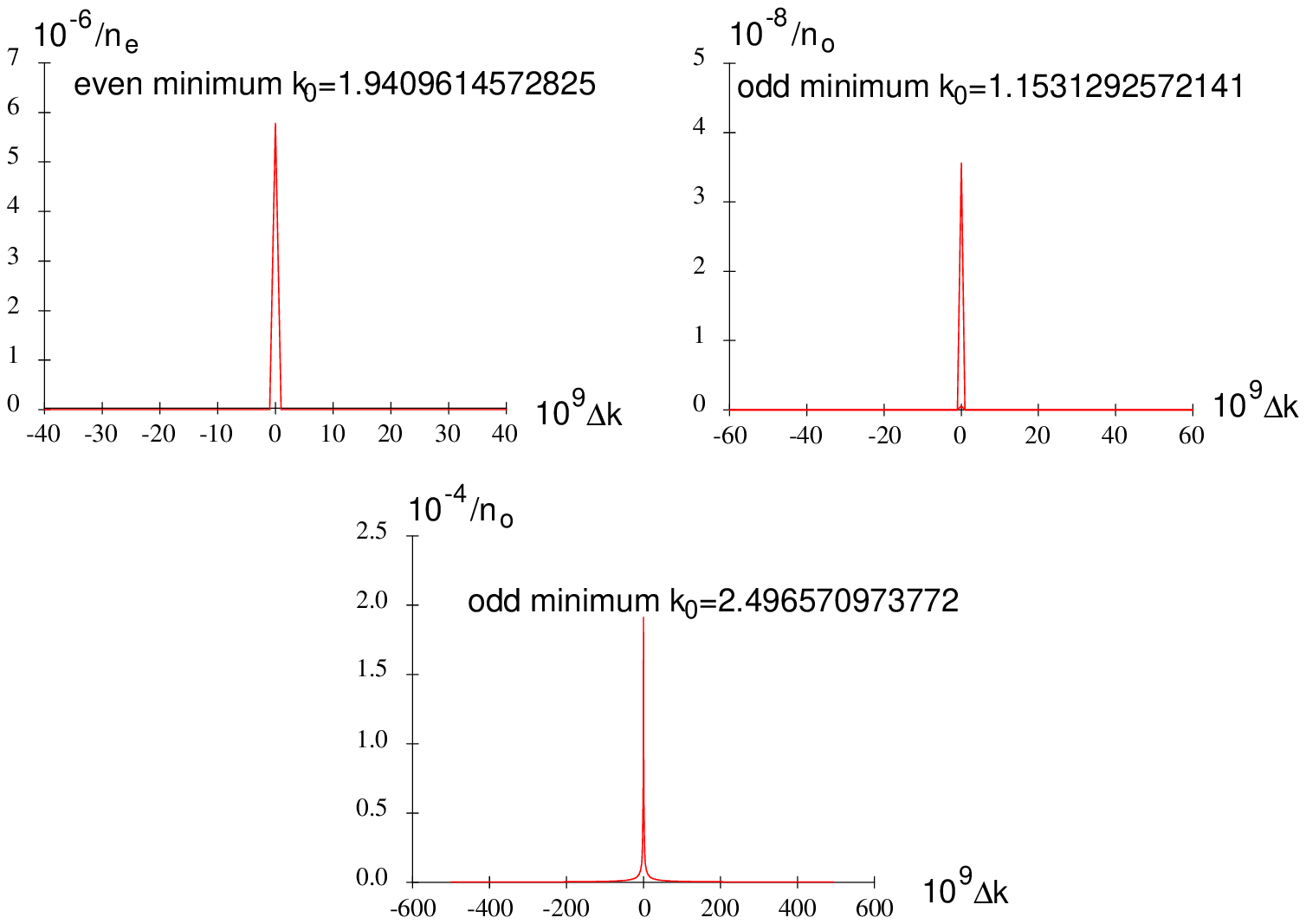}
	\caption{Normalizations of 
eq.(\ref{funcs}) around the minima at $\ds k=k_0$, with $\ds \Delta k=k-k_0, see text.$}
\label{fig1}
\end{figure}

\noindent The number of poles is directly related to the width and height of the barrier. 
Very few minima and consequently very few poles show up below the barrier energy for lighter nuclei, 
as compared to the $\ds \alpha$ decay
process from heavier nuclei for which the repulsive barriers are wider and taller.\ci{g2}

The normalization factors around their minima can be cast in the form\ci{g1}

\bea\label{pole_1}
k^2~n_{e,o}^2&\approx&~\lambda_{j,(e,o)}~(k^2-{k_{j,(e,o)}}^2)^2
+\beta_{j,(e,o)}
\eea
\noindent where {\sl j} enumerates the zeros of $\ds n_e, n_o$.

The  poles 
appear in pairs located symmetrically above and below the real 
momentum axis\ci{g2}. The stationary wave functions of eqs.(\ref{even},\ref{odd})
consist of incoming and outgoing parts that have poles in the upper and lower 
halves of the complex energy plane respectively\ci{baz}.

The minima of the normalization factors are extremely sharp. It was necessary to apply a
multiple precision numerical package to locate them \ci{fmlib}
with a maximum of 50 decimal digits.
The imaginary parts of the poles are orders of magnitude smaller 
than the real parts.\footnote{The numerical results of \ci{g2}
were not obtained with the the multiple precision package FMLIB. 
The errors incurred can be estimated to be at most 15\%.
The product of the even and odd amplitudes at a minimum is now found to be exactly equal to one,
 and not approximately so.}

Eq.(\ref{expansion}) can be performed now by contour integration
in the complex energy plane, closing the contour from below to insure 
convergence.
The negative imaginary part of each pole induces a decaying exponential
when the dominant time dependence is due to the phase of the unperturbed wave function.
The perturbation can induce both decaying and growing behaviors. 
In \ci{g1},\ci{g2} we were interested in the acceleration of the decay. 
Here, we seek an enhancement of the
 rate of penetration into the well region, for which a different parameter space will be needed.

After performing the complex energy plane contour integration, 
the wave function in the region between the barriers becomes

\bea\label{psidev}
\Phi&=&\Phi_{int}+\Phi_{ext}\nono
\Phi_{ext}&=&\xi_{1,e}+\xi_{2,o}+\xi_{3,o}+\xi_{4,e}\nono
\Phi_{int}&=&\sum_{i=e,o}(\Xi_{1,i}+\Xi_{2,i}+\Xi_{3,i}+\Xi_{4,i})
\eea

\noindent The splitting into an external and internal contribution is necessary
because the outer part of the wave function vanishes 
at a pole. The treatment of the contour
integration has to be modified accordingly.

The rather involved expressions of the wave functions of 
eqs.(\ref{psidev}) are spelled out in appendix B.

Using the large argument expansion of the hypergeometric functions\ci{abram} in eqs.(\ref{xi1}-\ref{xi4}) and
eqs.(\ref{xi1af}-\ref{xi4af}) of Appendix B, the 
long time $\ds t\ra \infty$ behavior of eq.(\ref{psidev}) in the region of the well $\ds |x|<x_0$ becomes

\bea\label{longtime}
\Phi(x,t)&\approx&\sum_j\frac{C_j(x)}{\sqrt{t}}~e^{F_j(t)}\nono
F_j(t)&=&\frac{\mu k_j^R~k_j^I~Y(t)}{m~|q_0|}-\frac{k_j^R~k_j^I~t}{m}
\eea

\noindent where we have used the connection between the even and odd
amplitudes at the poles\ci{g2}

\bea\label{eq1}
n_o(k_{o,j})^2\beta_{e,j}^2&=&k_{e,j}^2\nono
n_e(k_{e,j})^2\beta_{o,j}^2&=&k_{o,j}^2
\eea

\noindent $k_j^R,~k_j^I~$ denote the real and imaginary parts
of the poles, $\ds |q_0|=\sqrt{2~m~V_0}$, and $\ds C_j(x)$ is a prefactor
that depends on the specific features of the initial wave packet.

Inspection of eq.(\ref{longtime}) reveals that the perturbation can
either accelerate the tunneling process of slow it down depending on the
sign of $\ds F(t)$. This is also true for the assisted tunneling corresponding to
 $\ds \alpha $ decay. 
 In \ci{g1},\ci{g2} we focused on assisted 
decay for which $\ds F(t)$ was demanded to be negative.

It follows from eq.(\ref{longtime}) that the perturbation will assist the fusion process when

\bea\label{condition}
\frac{\mu~Y(t)}{|q_0|}>t
\eea
\noindent independently of the pole structure.

To fulfill eq.(\ref{condition}), $\ds Y(t)$ has to be positive
definite. 

Using eqs.(\ref{th},\ref{ampl}),$\ds G(t)=sin(\Omega~t)$, with $\ds \Omega \ne 0$,

\bea\label{Y}
Y(t)=\frac{(\Omega~t-sin(\Omega~t))}{\Omega^2}
\eea

\noindent$Y(t) >0$ is satisfied for $\ds \Omega~t>>1$.

For long times, eq.(\ref{condition}) becomes

\bea\label{cond1}
\Omega< \frac{\mu}{|q_0|}
\eea

It is now possible now to estimate the parameters needed for assisted fusion.

Consider low density and temperature totally ionized oxygen nuclei 
approximated as an ideal gas. For a temperature of $\ds 100 ^0K$ the velocity of
the oxygen nuclei is $\ds~v\approx 400\frac{m}{sec}$. The inter nuclei distance
at $\ds P = 0.1 Pa$ is $\ds \approx 2400 \mathring{A}$. The average time 
between collisions is $\delta t\approx 6~10^{-10} sec$.  If we take $\ds \Omega~\delta t=10$,  
eq.(\ref{cond1}) becomes

\bea\label{cond2}
\mu>10~\frac{|q_0|}{t}
\eea

\noindent Inserting the fusion parameters used in figure 1 into eq.(\ref{cond2}) we find

\bea\label{cond3}
\mu> 600 \frac{MeV}{cm}
\eea
\noindent or equivalently an electric field amplitude in eq.(\ref{th}) of around $\ds |E_0|
\approx 8~10^7 \frac{Volt}{cm}$. 

This electric field appears quite large but not unreachable, 
especially in light of the fact that the corresponding frequency of eq.(\ref{cond1})
is $\ds \Omega\approx 17 Ghz$ and the perturbation can be applied
in ultrashort pulses of nanosecond duration.

For angular frequencies smaller than the value prescribed by eq.(\ref{cond1}), the tunneling into the well
of a packet located far away outside the well, will increase
as compared to the unassisted case. 

Undoubtedly, the actual fusion problem is more complicated than the simplified
model of a packet impinging on a well between barriers, especially because of nuclear structure aspects. However,
the analysis of tunneling provided here suggests that the enhancement
of fusion by means of external time dependent agents is possible.

\begin{acknowledgments}

\noindent I would like to express my gratitude to Prof. David Smith
of Loyola University for his patient guidance in implementing the FMLIB multiple
precision package.
\end{acknowledgments}

\section{\sl Appendix A: Amplitudes of the unperturbed wave functions}

The amplitudes of the unperturbed wave functions of eqs.(\ref{even},\ref{odd}) are

\bea\label{amp1}
A_1&=&-\frac{1}{2\kappa~e_1}( q s_1-\kappa~c_1)\nono
B_1&=&-\frac{e_1}{2\kappa}( q s_1+\kappa~c_1)\nono
A_2&=&\frac{1}{2\kappa~e_1}( q c_1+\kappa~s_1)\nono
B_2&=&-\frac{e_1}{2\kappa}( q c_1-\kappa~s_1)
\eea

\bea\label{neven}
((n_e(k))^2&=&(C_1(k)^2+D_1(k)^2)\nono
C_1&=&\frac{1}{2~e_1~e_2~q~k}(e_2^2~q~\kappa
~s_1 s_2-e_2^2\kappa^2~c_1~s_2\nono
&+&e_1^2\kappa^2~c_1~s_2+e_1^2~q~\kappa~s_1~s_2\nono
&-&e_2^2~q~k~s_1~c_2+e_2^2~\kappa~k~c_1~c_2\nono
&+&e_1^2~\kappa~k~c_1~c_2+e_1^2~q~k~s_1~c_2)\nono
D_1&=&\frac{1}{2~e_1~e_2~q~k}(-e_2^2~q~\kappa
~s_1 c_2+e_2^2\kappa^2~c_1~c_2\nono
&-&e_1^2\kappa^2~c_1~c_2-e_1^2~q~\kappa~s_1~c_2\nono
&-&e_2^2~q~k~s_1~s_2+e_2^2~\kappa~k~c_1~s_2\nono
&+&e_1^2~\kappa~k~c_1~s_2+e_1^2~q~k~s_1~s_2)
\eea

\bea\label{nodd}
((n_o(k))^2&=&(C_2(k)^2+D_2(k)^2)\nono
C_2&=&\frac{1}{2~e_1~e_2~q~k}(-e_2^2~q~\kappa
~c_1 s_2-e_2^2\kappa^2~s_1~s_2\nono
&+&e_1^2\kappa^2~s_1~s_2-e_1^2~q~\kappa~c_1~s_2\nono
&+&e_2^2~q~k~c_1~c_2+e_2^2~\kappa~k~s_1~c_2\nono
&+&e_1^2~\kappa~k~s_1~c_2-e_1^2~q~k~c_1~c_2)\nono
D_2&=&\frac{1}{2~e_1~e_2~q~k}(e_2^2~q~\kappa
~c_1 c_2+e_2^2\kappa^2~s_1~c_2\nono
&-&e_1^2\kappa^2~s_1~c_2+e_1^2~q~\kappa~c_1~c_2\nono
&+&e_2^2~q~k~c_1~s_2+e_2^2~\kappa~k~s_1~s_2\nono
&+&e_1^2~\kappa~k~s_1~s_2-e_1^2~q~k~c_1~s_2)
\eea

\bea\label{ne_1}
\kappa&=&\sqrt{2~m~\gamma-k^2}\nono
q&=&\sqrt{2~m~(k^2+V_0)}\nono
e_1&=&e^{\kappa~x_0}\nono
e_2&=&e^{\kappa~d}\nono
c_2&=&cos(k~d)\nono
s_2&=&sin(k~d)\nono
c_1&=&cos(q~x_0)\nono
s_1&=&sin(q~x_0)\nono
\eea

\section{\sl Appendix B: Wave function in the region between the barriers}

The even and odd amplitudes of eq.(\ref{aeao}) are written as

\bea\label{aeaop}
A_{e,o}&=&\frac{{\bf A}_{e,o}}{n_{e,o}}\nono
&=&\frac{{\bf A}_{ext}^{e,o}+{\bf A}_{int}^{e,o}}{n_{e,o}}\nono
{\bf A}_{ext}^{e,o}&=&=\int_{-\infty}^{\infty}\Theta(|x|-d)
\Psi_{q_0,r_0}\chi_{e,o}~dx\nono
{\bf A}_{int}^{e,o}&=&=\int_{-\infty}^{\infty}\Theta(d-|x|)
\Psi_{q_0,r_0}\chi_{e,o}~dx
\eea

\noindent Splitting of the contributions of the inner and outer regions
is needed because the outer part of the wave functions vanishes at the poles.
The finite part of the time dependent wave function for the outer regions 
has to be evaluated by a slightly different method than the
contribution of the inner part.

For the sake of exemplification, we consider an initial Gaussian wave packet 
carrying mean momentum $\ds q_0$ and centered at $\ds r_0$,

\bea\label{gauss}
\Psi_{q_0,r_0}(x,t=0)&=&e^{\phi}\nono
\phi&=&i~q_0~(x-r_0)-\frac{(x-x_0)^2}{\Delta^2}
\eea

Inserting eq.(\ref{aeaop}) into the wave function of 
eqs.(\ref{psidev}) for the region between the barriers, the contour integration around the
poles yield

\bea\label{xi1}
\Xi_{1,e}&=&\sum_j~\sum_{n=0}^{\infty}~Q_{ee}\frac{(-X_{j,e}^2)^n}
{(n!)^2}\nono
Q_{ee}&=&\sqrt{\pi}~{\bf A}^{int}_{e}~cos(k_{j,e}~x)~e^{w_e}\nono
&&\frac{~k_{j,e}}{2~\sqrt{2\beta_{j,e}\lambda_{j,e}}}\nono
\Xi_{1,o}&=&\sum_j~\sum_{n=0}^{\infty}~Q_{eo}\frac{(-X_{j,o}^2)^n}
{(n!)^2~(2~n+1)~(n+1)}\nono
Q_{eo}&=&-\sqrt{\pi}{\bf A}_o cos(k_{j,o}~x) e^{w_o}\mu^2~Y(t)^2 q_{j,o}^2\nono
&&\frac{k_{j,o}}{4 \sqrt{2\beta_{j,o}\lambda_{j,o}}~m^2~n_e^4(k_{j,o})}
\eea

\noindent where

\bea\label{vars}
X_{j,(e,o)}&=&\mu~q_{j,(e,o)}~Y(t)\nono
&&\frac{k_{j,(e,o)}}{2~m~n_o(k_{j,(e,0)})\sqrt{\beta_{j,(e,o)}}}\nono
w_{e,o}&=&-k_{j,(e,o)}^2(\frac{\Delta^2}{4}+\frac{i~t}{2~m})\nono
q_{j,(e,o)}&=&\sqrt{(k_{j,(e,o)})^2+2~m~V_0}
\eea

\bea\label{xi2}
\Xi_{2,e}&=&\sum_j~\sum_{n=0}^{\infty}~{\tilde Q}_
{ee}\frac{(-X_{j,e}^2)^n}
{(n!)^2~(2~n+1)}\nono
{\tilde Q}_{ee}&=&\mu~Y(t)~q_{j,e}
\sqrt{\pi}{\bf A}_o~cos(k_{j,e}~x)~e^{w_e}\nono
&&\frac{k_{j,e}}{2~m~n_o^2(k_{j,e})~\sqrt{2\beta_{j,e}\lambda_{j,e}}}\nono
\Xi_{2,o}&=&\sum_j~\sum_{n=0}^{\infty}~{\tilde Q}_{eo}\frac{(-X_{j,o}^2)^n}{(n!)^2~(2~n+1)}\nono
{\tilde Q}_{eo}&=&\mu~Y(t)~q_{j,o}
\sqrt{\pi}{\bf A}^{int}_{o})~e^{w_o}\nono
&&\frac{~k_{j,o} cos(k_{j,o})}{2~m~n_e^2(k_{j,o})~\sqrt{2\beta_{j,o}\lambda_{j,o}}}
\eea

\bea\label{xi3}
\Xi_{3,e}&=&\sum_j~\sum_{n=0}^{\infty} Q_{oe}\frac{(-X_{j,e}^2)^n}
{(n!)^2(2n+1)(n+1)}\nono
Q_{oe}&=&-\mu^2~Y(t)^2~q_{j,e}^2 \sqrt{\pi}{\bf A}^{int}_{o} sin(k_{j,e}~x) e^{w_e}\nono
&&\frac{k_{j,e}}{4 \sqrt{2\beta_{j,e}\lambda_{j,e}} m^2 n_o^4(k_{j,e})}\nono
\Xi_{3,o}&=&\sum_j~\sum_{n=0}^{\infty}~Q_{oo}\frac{(-X_{j,o}^2)^n}{(n!)^2}\nono
Q_{oo}&=&\sqrt{\pi}{\bf A}_o~sin(k_{j,o}~x)~e^{w_o}\nono
&&\frac{k_{j,o}}{2~\sqrt{2\beta_{j,o}\lambda_{j,o}}}
\eea

\bea\label{xi4}
\Xi_{4,e}&=&\sum_j~\sum_{n=0}^{\infty}~{\tilde Q}_
{oe}\frac{(-X_{j,e}^2)^n}
{(n!)^2~(2~n+1)}\nono
{\tilde Q}_{oe}&=&-\mu~Y(t)~q_{j,e}~\sqrt{\pi}{\bf A}^{int}_e~sin(k_{j,e}~x)~e^{w_e}\nono
&&\frac{k_{j,e}~}{2~m~n_o^2(k_{j,e})~\sqrt{2\beta_{j,e}\lambda_{j,e}}}\nono
\Xi_{4,o}&=&\sum_j~\sum_{n=0}^{\infty}~{\tilde Q}_
{oo}\frac{(-X_{j,o}^2)^n}{(n!)^2~(2~n+1)}\nono
{\tilde Q}_{oo}&=&-\mu~Y(t)~q_{j,o}\sqrt{\pi}{\bf A}_e~sin(k_{j,o}~x)~e^{w_o}\nono
&&\frac{~k_{j,o}}{2~m~n_e^2(k_{j,o})~\sqrt{2\beta_{j,o}\lambda_{j,o}}}
\eea

\bea\label{xi1af}
\xi_{1,e}&=&\sum_j~\sum_{n=0}^{\infty}~Q^{ext}_{ee}\frac{(-X_{j,e}^2)^n (4n)!}
{(2n!)^3}\nono
Q^{ext}_{ee}&=&\frac{\sqrt{\pi}{\bf A}^{ext}_e~cos(k_{j,e}~x)~e^{w_e}}
{2~\sqrt{\lambda_{j,e}}}
\eea

\bea\label{xi2af}
\xi_{2,o}&=&\sum_j~\sum_{n=0}^{\infty}~{\tilde Q}^{ext}_
{eo}\frac{(-X_{j,o}^2)^n~(4~n)!}
{(2~n!)^2~(2~n+1)}\nono
{\tilde Q}^{ext}_{eo}&=&\frac{\mu~Y(t)~q_{j,o}
\sqrt{\pi}{\bf A}^{ext}_o~cos(k_{j,o}~x)~e^{w_o}}
{2~m~n_e^2(k_{j,o})~\sqrt{\lambda_{j,o}}}
\eea

\bea\label{xi3af}
\xi_{3,o}&=&\sum_j~\sum_{n=0}^{\infty}~Q^{ext}_{oo}\frac{(-X_{j,o}^2)^n (4n)!}
{(2n!)^3}\nono
Q^{ext}_{oo}&=&\frac{\sqrt{\pi}{\bf A}^{ext}_o~sin(k_{j,o}~x)~e^{w_o}}
{2~\sqrt{\lambda_{j,o}}}
\eea

\bea\label{xi4af}
\xi_{4,e}&=&\sum_j~\sum_{n=0}^{\infty}~{\tilde Q}^{ext}_
{oe}\frac{(-X_{j,e}^2)^n~(4~n)!}
{(2~n!)^2~(2~n+1)}\nono
{\tilde Q}^{ext}_{oe}&=&-\frac{\mu~Y(t)~q_{j,e}
\sqrt{\pi}{ A}_{ext,e}~sin(k_{j,e}~x)~e^{w_e}}
{2~m~n_o^2(k_{j,e})~\sqrt{\lambda_{j,e}}}
\eea

\noindent where $\ds k_{j,e}$ is the complex momentum at the pole 
number {\sl j} of the even unperturbed wave function, Y(t) is defined
in eq.(\ref{ampl}), and  $\ds \lambda$ and $\ds\beta$ correspond to the
expansion around a pole of eq.(\ref{pole_1}).
The wave functions of eqs.(\ref{xi1}-\ref{xi4}) and
eqs.(\ref{xi1af}-\ref{xi4af}), 
can be expressed in terms of standard hypergeometric, Struve, and Bessel 
functions.\ci{abram}

\end{document}